# Arbitrary control of the flow of light using pseudomagnetic fields in photonic crystals at telecommunication wavelengths

Pan Hu[1], Lu Sun[1*], Ce Chen[1], Jingchi Li[1], Xiong Ni[1], Xintao He[2], Jianwen Dong[2*], and Yikai Su[1*]

[1]State Key Lab of Advanced Optical Communication Systems and Networks, Department of Electronic Engineering, Shanghai Jiao Tong University, Shanghai 200240, China

[2]State Key Laboratory of Optoelectronic Materials and Technologies, School of Physics, Sun Yat-Sen University, 510275 Guangzhou, China

*Corresponding author: sunlu@sjtu.edu.cn, dongjwen@mail.sysu.edu.cn, yikaisu@sjtu.edu.cn

**Abstract**

In photonics, the idea of controlling light in a similar way that magnetic fields control electrons has always been attractive. It can be realized by synthesizing pseudomagnetic fields (PMFs) in photonic crystals (PhCs). Previous works mainly focus on the Landau levels and the robust transport of the chiral states. More versatile control over light using complex nonuniform PMFs such as the flexible splitting and routing of light has been elusive, which hinders their application in practical photonic integrated circuits. Here we propose an universal and systematic methodology to design nonuniform PMFs and arbitrarily control the flow of light in silicon PhCs at telecommunication wavelengths. As proofs of concept, a low-loss S-bend and a highly efficient 50:50 power splitter based on PMFs are experimentally demonstrated. A high-speed data transmission experiment is performed on these devices to prove their applicability in real communication systems. The proposed method offers a new paradigm for the exploration of fundamental physics and the development of novel nanophotonic devices.

**Introduction**

In condensed-matter physics, many intriguing phenomena can be observed for electrons under external magnetic fields, e.g., the quantum Hall effect and the Landau levels [1-12]. These phenomena play a pivotal role in shaping our understanding of the topological phases of matter [13-16]. However, neutral particles, such as photons and cold atoms, barely interact with magnetic fields. Although the magneto-optic effects are strong enough at microwave frequencies to effectively mediate the interactions between photons and magnetic fields [17, 18], the magneto-optic coefficients of natural materials are too weak at optical frequencies, making the magnetic-field-associated effects inaccessible in on-chip nanophotonic systems. Pseudomagnetic fields (PMFs), also known as artificial gauge fields, have been proposed to mimic the behaviours of electrons under magnetic fields in classical wave systems [19-40]. Recently, Landau levels and Landau rainbow were realized in two-dimensional (2D) photonic crystals (PhCs) by inducing PMFs via synthetic strain or breaking local spatial inversion symmetry in each unit cell [41-44]. Spin- and valley-Hall effects were also observed in photonic Dirac waveguides and cavities [45-48]. However, these works focus on the Landau levels and the behaviours of the related chiral states similar to those of the topological edge states in topological PhCs. A systematic method based on PMFs for the more flexible control over light is still missing. It hinders their application in practical photonic integrated circuits where more complicated functions such as arbitrary splitting and routing of light are necessary.

Here we go beyond the Landau levels and the chiral states and propose an universal design method of PMFs which can realize the arbitrary control of the flow of light in silicon PhCs at telecommunication wavelengths. By changing the local spatial-inversion-symmetry-breaking strength at every point in PhCs, effective PMFs with arbitrary distributions can be designed at will and introduced into dielectric photonic systems without breaking the real time-reversal symmetry. The supported modes and the propagation of light in the PhCs with the synthetic PMFs are theoretically and experimentally investigated. As examples of routing and splitting light using PMFs, a S-bend with a low insertion loss (IL) of <1.83 dB and a 50:50 power splitter with a low excess loss (EL) of <2.11 dB and imbalance of <±0.5 dB are proposed and experimentally demonstrated. To prove the applicability of these functional devices based on PMFs in practical on-chip photonic systems, we conduct a high-speed data transmission experiment with a 140-Gb/s four-level pulse amplitude modulation (PAM-4) signal per channel. To the best of our knowledge, this is the first time that a systematic method is proposed to synthesize PMFs for arbitrarily controlling the flow of light and realizing various functional devices for on-chip optical communications. Our findings pave the way toward developing a new class of nanophotonic devices with PMFs, which may find applications in many fields including optical communications, optical computing, quantum information processing and beyond.

**Results**

**Realization of pseudomagnetic fields in photonic crystals**

Figure 1A shows a schematic of the PhC with a honeycomb lattice implemented on the silicon-on-insulator (SOI) platform. The unit cell is illustrated in the red dashed rhombic boxes and contains two inverted equilateral triangular holes with side lengths $d_1$ and $d_2$. The lattice constant is $a_0$. When the spatial inversion symmetry is unbroken ($d_1 = d_2$), the PhC lattice exhibits a $C_{6v}$ symmetry, featuring a Dirac cone at the $K$ and $K'$ points in the momentum space, as shown by the band diagram of the transverse electric (TE) modes of the PhC in Fig. 1A. By using the plane wave expansion (PWE) method and applying the $\vec{k} \cdot \vec{p}$ approximation [49-54], the two lowest bands can be described analytically by an effective Hamiltonian near the $K/K'$ points: $H_{K/K'} = \pm(vk_x\sigma_x + vk_y\sigma_y)$, where $\pm$ represents the $K$ ($K'$) valley pseudospin, $\sigma_{x,y,z}$ denotes the Pauli matrices, $v$ is the group velocity, and ($k_x$, $k_y$) is the reciprocal vector with respect to the degeneracy points $K$ and $K'$ (see supplementary text S1 for more details). By introducing asymmetry between the sizes of the two holes within the unit cell ($d_1 \neq d_2$), the spatial inversion symmetry is broken and the lattice symmetry is reduced from $C_{6v}$ to $C_{3v}$. An effective mass term $m$ is introduced to the Dirac cone, which lifts the Dirac degeneracy and opens a bandgap of $\Delta\omega = 2|m|$ (see supplementary text S2 for more details). If the mass term varies with position, the effective

Hamiltonian becomes:

$$H_{K/K'} = \pm[vk_x\sigma_x + vk_y\sigma_y + m(\vec{r})\sigma_z] \tag{1}$$

where $m(\vec{r})$ is the position-dependent mass term. Considering a Hamiltonian describing the interaction between the Dirac quasiparticles and the synthetic gauge field $A_z$, it should take the form $H_{K/K'} = \pm[vk_x\sigma_x + vk_y\sigma_y + p_z\sigma_z]$, where $p_z = k_z + A_z$ corresponds to the canonical momentum in the presence of the magnetic field. By comparing it with Eq. 1 and noticing that $k_z = 0$ in the case of 2D PhCs discussed here, the effective mass term is equivalent to a magnetic vector potential, i. e., $A_z = m(\vec{r})$. For a linearly varying mass term $m(x) = ax$, it corresponds to a PMF $B_y = -a$, which gives rise to the celebrated Landau levels. Most importantly, the fields of the chiral states of the Landau levels will be confined to the vicinity of the area that satisfies $p_z = 0$ or equivalently $A_z = 0$. Different from topological edge states, the Landau level states are propagative bulk states, which are also topologically protected. Figure 1B shows the PhC structure, the band diagram and the Bloch mode profile of the zeroth-order Landau level in one period along the *y* direction. In the *x* direction, the side length of the triangular holes $d_1$ ($d_2$) increases (decreases) linearly, as shown in the supercell encircled by the red dashed boxes. The red and green dotted lines in the band diagram represent the chiral states of the zeroth-order Landau level at the *K* and *K'* points, respectively. Since they are well separated from the other Landau levels (black dotted lines) and therefore their states are easy to be excited, hereafter we choose to work with the zeroth-order Landau level to construct various nanophotonic devices in silicon PhCs. The states of the zeroth-order Landau level are approximately linearly dispersive and the dispersion relationship can be expressed as $\omega_{K/K'} = \pm\text{sgn}(B_y)vk_y$ [30, 44], where $\omega_{K/K'}$ is the frequency of the mode in the *K* (*K'*) valley, and $\text{sgn}(B_y)$ is the sign of the synthetic magnetic field. On the bottom of Fig. 1B shows the $H_z$ field distribution of the state corresponding to the blue star point on the green dotted line in the band diagram. Obviously, the field is highly localized near $x = 0$ (purple shaded area in Fig. 1B) where the condition $A_z = 0$ is satisfied. The expression of the state can be derived

by solving Eq. 1 (see supplementary text S1 for details):

$$\psi(x,y) = \frac{1}{\sqrt{2}}\begin{pmatrix}1\\i\end{pmatrix}e^{ik_y y}e^{-\frac{1}{v}\int m(x)dx} = \frac{1}{\sqrt{2}}\begin{pmatrix}1\\i\end{pmatrix}e^{\frac{-a}{2v}x^2}e^{ik_y y} \quad (2)$$

The state shows the behavior of a propagating wave along the $y$ direction and the characteristics of a Gaussian wave along the $x$ direction, indicating the existence of a propagative bulk state near $x = 0$. Moreover, the field distribution of the state is directly related to the effective mass term, i. e., $\psi(x,y) \propto e^{-\frac{1}{v}\int m(x)dx}$, which means the confinement of light near $x = 0$ is guaranteed as long as $m(x)$ is an odd function such as $m(x) = ax^{1/3}$ and $m(x) = ax^3$, and can be tuned by choosing different functions for $m(x)$ (see supplementary text S3 for details). In Fig. 1C, we show another example with a linearly varying mass term $m(y) = by$, which corresponds to a PMF $B_x = b$. The side length of the triangular holes $d_1$ ($d_2$) decreases (increases) linearly in the $y$ direction, as shown in the supercell in the red dashed boxes. The red dotted line in the band diagram represents the propagative bulk states of the zeroth-order Landau level. The profile of the mode corresponding to the blue star point is shown on the bottom right of Fig. 1C. As one can see, the field is well confined to the region near $y = 0$ (purple shaded area in Fig. 1C). More details about the propagative bulk states under such a PMF can be found in supplementary text S1.

By synthesizing uniform magnetic fields, one can realize the Landau levels and the straight waveguiding of light in the $x$ and $y$ directions. However, more complicated manipulation of light wave is required for constructing photonic integrated circuits for real applications. We go beyond the Landau levels and design complex nonuniform PMFs to flexibly control the flow of light on a chip, as shown in Fig. 1D. By introducing a nonuniform PMF varying in both the $x$ and $y$ directions, the deflection of light can be realized and exploited to build a S-bend in a PhC platform, as indicated by the purple shaded stripe on the left of Fig. 1D. Different from the sharp bends that exploit the spin- and valley-Hall effects [45, 49], the S-bend is solely determined by the predefined PMF and therefore provides more flexibility in designing the optical path. Furthermore, the field distribution of the propagative bulk state can be engineered by judiciously designing the PMF through the relationship $\psi(x,y) \propto e^{-\frac{1}{v}\int m(x,y)dx}$. In other words, once the light path or the field distribution function $\psi(x,y)$ is given, one can derive the corresponding effective mass distribution $m(x,y)$ by taking the first-order derivative of $\psi(x,y)$ with respect to the spatial coordinate $x$. This method can be employed to build a power splitter with a splitting ratio of 50:50 in a PhC platform, as schematically illustrated on the right side of Fig. 1D. The light paths in the PhC are shaded in purple as a guide to the eye. The design and the implementation of the PMFs for these two devices will be detailed in the next section. Most importantly, we argue that the methodology presented here is universal and systematic. A new family of nanophotonic devices can be developed by designing and synthesizing nonuniform PMFs with this method.

**Device design and experimental demonstration**

We start with the straightforward-propagating states of the zeroth-order Landau levels under the uniform PMFs $m(x) = ax$ and $m(y) = by$. The design details, simulation results and experimental results are given in supplementary text S4. The ILs of the two devices are lower than 1.32 dB and 0.8 dB respectively in the wavelength range of 1520-1580 nm, manifesting the capability of on-chip waveguiding for these two kinds of PMFs. Then we move on to the more general cases where the more sophisticated nonuniform PMFs are designed and implemented for various basic functional devices such as a S-bend and a power splitter. Figure 2A shows a schematic of a PhC-based S-bend operating at telecommunication wavelengths which is built on a SOI platform with a 220-nm-thick top silicon layer, a 3-μm-thick buried oxide layer and a 1-μm-thick silica upper cladding layer. The PMF is designed to be $m(x,y) = cx+d$ in Region I, $m(x,y) = ax+by$ ($b/a = -10$) in Region II, and $m(x,y) = cx-d$ in Region III. It leads to a zero magnetic vector potential $A_z = 0$ on the line $x = -d/c$ in Region I,

$y = -ax/b$ in Region II and $x = d/c$ in Region III, which defines the propagation path of light in the PhC as indicated by the purple shaded stripe in Fig. 2A. The lattice constant of the PhC is $a_0 = 450$ nm. The side lengths of the triangular holes $d_1$ and $d_2$ are 225 nm on the propagation path line and change linearly with a variation step of 14.7 nm ($\Delta d_1 = -\Delta d_2 = 14.7$ nm) along the direction perpendicular to the line. The light is coupled into and out of the chip by grating couplers, as illustrated in Fig. 2A. The PhC structure is connected to the grating couplers via two 3-µm-wide silicon stripe waveguides whose fundamental TE mode excites the zeroth-order propagative bulk state in the PhC. To prove that the bending of light based on the nonuniform PMF works as well as the straight waveguiding of the Landau level state under the uniform PMF, we used the three-dimensional finite-difference time-domain (3D FDTD) methods to simulate the propagation of light in both the straight waveguide and the S-bend (The simulation details can be found in Materials and Methods). Figure 2B presents the simulated propagation profiles for the two devices at the wavelength of 1550 nm. It is evident that the light propagates along a S-shaped trajectory predefined by $A_z = 0$ in a nonuniform PMF just as in the case of a straight PhC waveguide, which is in good agreement with our theoretical prediction. These devices were also fabricated on the SOI platform using complementary metal-oxide-semiconductor (CMOS)-compatible process (The fabrication details can be found in Materials and Methods and supplementary text S5). The scanning electron microscopy (SEM) images of the fabricated devices are displayed in Fig. 2C, with the light propagation paths shaded in purple as a guide to the eye. The zoom-in images of the interface between the silicon stripe waveguide and the PhC, the area around $x = 0$ of the straight PhC waveguide and the area in the proximity of the propagation path in Region II of the S-bend are shown in the red, blue and orange dashed boxes on the bottom of Fig. 2C. Figure 2D shows the simulated and measured transmission spectra of both devices. The measured transmission spectra are all normalized to that of a pair of reference grating couplers fabricated on the same chip (see Materials and Methods and supplementary text S6 for the measurement details). The measured ILs of the straight waveguide and the S-bend are lower than 1.32 dB and 1.83 dB respectively in the wavelength range of 1520-1580 nm, which are in reasonable agreement with the simulated ILs of <2 dB (straight waveguide) and <2.27 dB (S-bend). We argue that the ILs mainly originate from the coupling losses at the interfaces between the silicon stripe waveguide and the PhC structure while the propagation losses inside the PhC are negligible. It is noteworthy that the S-bend structure introduces almost no extra losses as compared to the straight waveguide, confirming the feasibility of achieving low-loss nanophotonic devices with as-designed nonuniform PMFs.

Next, we present another example of a 50:50 power splitter to show how to design the nonuniform PMF according to the light field distribution $\psi(x, y)$. Figure 3A shows a schematic of the consecutive supercells of the power splitter along the propagation direction. The field of the propagative bulk state is initially localized around $x = 0$ and gradually becomes concentrated on two well-separated positions in a supercell, as indicated by the red Gaussian-like wave packets in Fig. 3A. The separation between the two maxima of the field grows linearly with propagation distance, implementing the splitting of light during propagation. The relationship between the light field and the PMF can be described by $\psi(x, y) \propto e^{-\frac{1}{v}\int m(x,y)dx}$ (see supplementary text S1 for details). By taking the first-order derivatives of $\psi(x, y)$ with respect to the spatial coordinate $x$, one can obtain the effective mass distribution $m(x, y)$ in every supercell, as shown on the bottom of Fig. 3A. The two zeros of the mass term separate slowly along the propagation direction. The whole structure of the power splitter is schematically illustrated in Fig. 3B where the propagation paths of light are shaded in purple. The lattice constant of the PhC is $a_0 = 450$ nm. The side lengths of the triangular holes $d_1$ and $d_2$ are 225 nm on the propagation paths where $A_z = 0$ and change gradually with a variation step of 6.9 nm away from the paths to mimic the nonuniform PMF described above (The relationship between the effective mass term and the sizes of the triangular holes can be found in supplementary text S2). Full-wave numerical simulations were carried out using the 3D FDTD methods to investigate the propagation of light in the power splitter. Figure 3C shows the propagation profile for the device at the wavelength of 1550 nm. The zeroth-order propagative bulk state is excited by the fundamental TE mode in a silicon stripe waveguide, split equally into two beams in the PhC, and output from two silicon stripe waveguides. Again, the proposed device was fabricated with CMOS-compatible process. Figure 3D presents the SEM images of the fabricated device, with the purple shaded area indicating the light propagation paths. The zoom-in images of the input port, the splitting area and the output ports are displayed in the red, blue and orange dashed boxes respectively on the bottom of Fig. 3D. The simulated and measured transmission spectra of the device are plotted in Fig. 3E. In the simulations, the imbalance of the power splitter is close to 0 dB and the EL is below 2.06 dB in the wavelength range of 1527-1573 nm. In the experiments, the measured imbalance and EL are lower than $\pm 0.5$ dB and 2.11 dB respectively in the same wavelength range, which is in good agreement with the simulation results. Moreover, the EL of the power splitter is comparable to the ILs of the straight waveguide and the S-bend shown in Fig. 2, proving that the EL mainly comes from the coupling loss between the silicon stripe waveguide and the PhC and the beam splitting process causes rare extra losses. More importantly, the arbitrary control of the flow of light with PMFs can be envisioned as both the light field distribution and the corresponding PMF can be designed at will using our method.

Since the intervalley coupling is negligible between the counterpropagating waves affiliated with different valleys, the transport of the propagative bulk states under the synthetic PMFs are topologically protected [41, 44]. Here we fabricated and measured different kinds of PhC devices (including the straight waveguides, the S-bend and the power splitter) with intentionally introduced defects. The experimental details are provided in supplementary text S7. No severe discrepancies were observed between the transmission spectra of the fabricated devices with and without defects. The experimental results show clear evidence that the performance of the PMF-based devices exhibits good robustness against small perturbations, which is very much desired in large-scale photonic integration and massive production.

**High-speed data transmission experiment**

To verify the feasibility of using the aforementioned devices in realistic on-chip optical communication systems, we performed a high-speed data transmission experiment based on the proposed S-bend and power splitter, as shown in Fig. 4. The experimental setup and the digital signal processing (DSP) flow charts for the transceiver are illustrated in Fig. 4, A and B, respectively (for details, see Materials and Methods and supplementary text S8). A Nyquist-shaped 70-GBaud PAM-4 signal is transmitted through each channel of the fabricated devices. The measured optical spectra of the signals before and after passing the devices are shown in Fig. 4, C and D. Figure 4E plots the bit error rates (BERs) for the channels of the S-bend and the splitter, which are all below the 7% hard-decision-forward error correction (HD-FEC) threshold of $3.8 \times 10^{-3}$. The recovered eye diagrams of the PAM-4 signals for different channels are presented in Fig. 4F. The demonstration provides unequivocal evidence that the basic functional devices built with PMFs in the telecom band can be used in real applications such as on-chip optical communications.

**Discussion**

In summary, we have proposed an universal and systematic methodology for designing and implementing PMFs to arbitrarily control the flow of light in silicon PhCs at telecommunication wavelengths. Different from the previous works that focus on the robust transport of the Landau level states (similar to the topological edge states in topological PhCs), we investigated the evolution of the propagative bulk states under nonuniform PMFs and realized diverse optical functions such as the bending and splitting of light on a chip with high flexibility. A S-bend with a low IL (<1.83 dB) and a 50:50 power splitter with a low EL (<2.11 dB) and imbalance (<±0.5 dB) were experimentally demonstrated. A high-speed data transmission experiment with 140-Gb/s PAM-4 signals was also carried out to verify the applicability of these devices in practical optical communication systems. Our work provides a new and highly flexible paradigm for controlling neutral particles with synthetic magnetic fields and designing novel nanophotonic devices with artificial gauge fields. The development of a number of PMF-based devices could be envisaged, such as power splitters with uneven splitting ratios, Mach-Zehnder interferometers, microloop resonators utilizing the spin- and valley-dependent transportation, and even wavelength multiplexing devices by introducing pseudoelectric fields simultaneously. It could serve as a design principle for a new family of nanophotonic devices that would help us understand the behaviours of photons under various PMFs and advance fields ranging from optical communications to optical computing and quantum information processing.

## Materials and Methods

### Numerical simulation

The band diagrams of the pristine PhC (Fig. 1A) and the supercell structures (Fig. 1, B and C) were calculated using the 2D simulations with the commercial software COMSOL Multiphysics. The effective index of the 220-nm-thick silicon slab was approximated as $n_{eff}$ = 2.8323 for all the 2D simulations. The propagation profiles and the transmission spectra for the straight waveguide (Fig. 2, B and D), the S-bend (Fig. 2, B and D) and the power splitter (Fig. 3, C and E) were simulated with the 3D FDTD methods using the commercial software Ansys Lumerical FDTD. The refractive index of silicon was taken as $n_{Si}$ = 3.48 for all the 3D simulations.

### Device fabrication

The fabrication started with a SOI wafer (220-nm-thick silicon layer on top of 3-μm-thick buried oxide layer). The wafer was first cleaned in ultrasound baths of acetone and isopropyl alcohol (IPA) and further cleaned using $O_2$ plasma asher. The patterns of the PhC structures and the stripe waveguides were defined on the photoresist (AR-P 6200.09) using electron beam lithography (EBL, Vistec EBPG 5200[+]). Then they were transferred onto the top silicon layer by inductively coupled plasma (ICP) dry etching (SPTS DRIE-I) with an etching depth of 220 nm. The above steps were repeated to define the grating couplers, but this time with an etching depth of 70 nm. After that, a

1-µm-thick silica cladding layer was deposited over the devices by plasma enhanced chemical vapor deposition (PECVD, Oxford Plasmalab System 100). The fabricated samples were inspected using the optical microscope and the SEM (Zeiss Ultra Plus). The detailed description of the fabrication procedure can be found in supplementary text S5.

**Optical characterization**

In the experiments, a tunable continuous wave (CW) laser (Santec TSL-770) and a photodetector (PD, Santec MPM-210) were employed to characterize the fabricated devices. The polarization of light from the tunable laser was first adjusted by a fiber polarization controller (PC). Then the TE-polarized light was coupled into and out of the chip through grating couplers. The coupling losses of the grating couplers are ~6 dB/port at the central wavelength of 1550 nm. An optical power meter was used for optical calibration. The wavelength of the laser was swept from 1500 nm to 1600 nm with the PD placed at the output end of the devices to measure the transmitted power at every wavelength. The transmission spectra of the devices were all normalized to that of the reference grating couplers, reflecting the spectral responses of these devices. The performance of the devices such as IL, EL and imbalance was assessed based on the transmission spectra. More details about the experimental setup and the measurement methods are provided in supplementary text S6.

**High-speed data transmission**

At the transmitter side, a 70-GBaud PAM-4 signal with a roll-off factor of 0.01 was resampled and sent to a digital-to-analog converter (DAC, Micram DAC4) with a sampling rate of 100 GSa/s. The output of the DAC was amplified by an electrical amplifier to drive a 25-GHz Mach-Zehnder modulator (MZM) biased at the quadrature point. A 10-dBm CW light output from the distributed feedback laser was injected into the MZM for the electrical-to-optical conversion. The modulated optical PAM-4 signal was amplified by an erbium-doped fiber amplifier (EDFA) before entering the on-chip devices. The receiver is comprised of an EDFA to compensate for the loss and a 70-

GHz PD for signal detection. The detected current was captured by an 80-GSa/s digital storage oscilloscope (DSO, LeCroy 36Zi-A). The receiver DSP contains resampling, matched filtering and synchronization. Then the transmission impairments were compensated by a least-mean square (LMS) algorithm-based linear feedforward equalizer (FFE), a post filter and a maximum-likelihood sequence decision (MLSD). Finally, the BERs were calculated for all the channels of the on-chip devices. More details about the experiment can be found in supplementary text S8.

**Acknowledgments**

We would like to thank the Center for Advanced Electronic Materials and Devices (AEMD) of Shanghai Jiao Tong University (SJTU) for its support in device fabrication. The work was supported in part by the National Key Research and Development Program of China (no. 2023YFB2905503) and the National Natural Science Foundation of China (nos. 62035016, 62475146, 62105200, and 62341508).

**Data and materials availability**

All data needed to evaluate the conclusions in the paper are present in the paper and/or the Supplementary Materials.

**Competing interests**

The authors declare that they have no competing interests.

**Figures and captions**

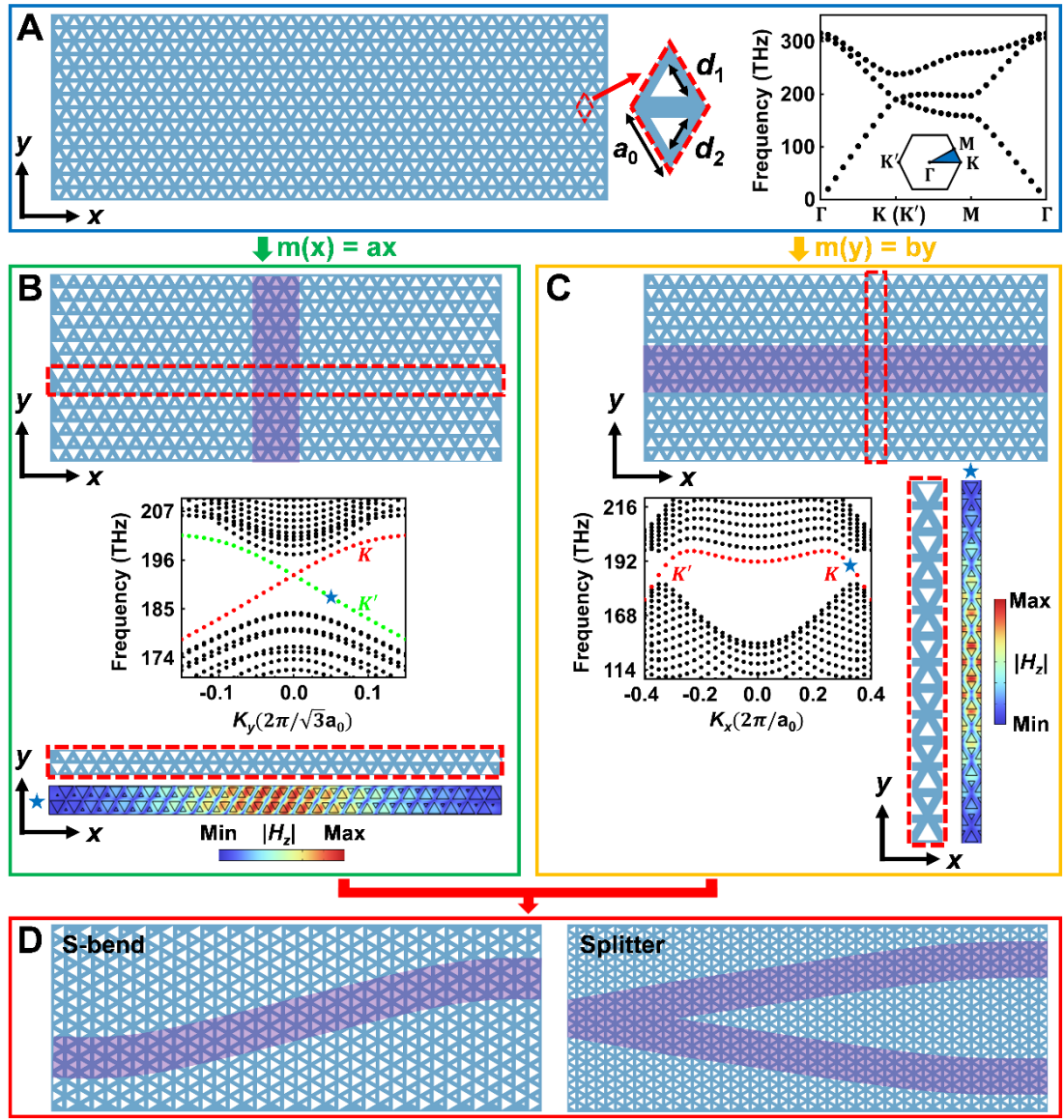

**Fig. 1. Realization of PMFs in PhCs.** (A) Schematic (left) and band diagram (right) of the pristine PhC with a honeycomb lattice. The rhombic unit cell is encircled by the red dashed line. The first Brillouin zone is shown in the inset of the band diagram. (B and C) Schematics of the PhCs (top), band diagrams of the supercells encircled by the red dashed line (middle/bottom left) and mode profiles corresponding to the blue star points in the band diagrams (bottom/bottom right) with inhomogeneous effective mass terms (B) $m(x) = ax$ and (C) $m(y) = by$. The propagation paths of light are shaded in purple. (D) Schematics of the S-bend (left) and the power splitter (right) based on PMFs, with the propagation paths of light shaded in purple.

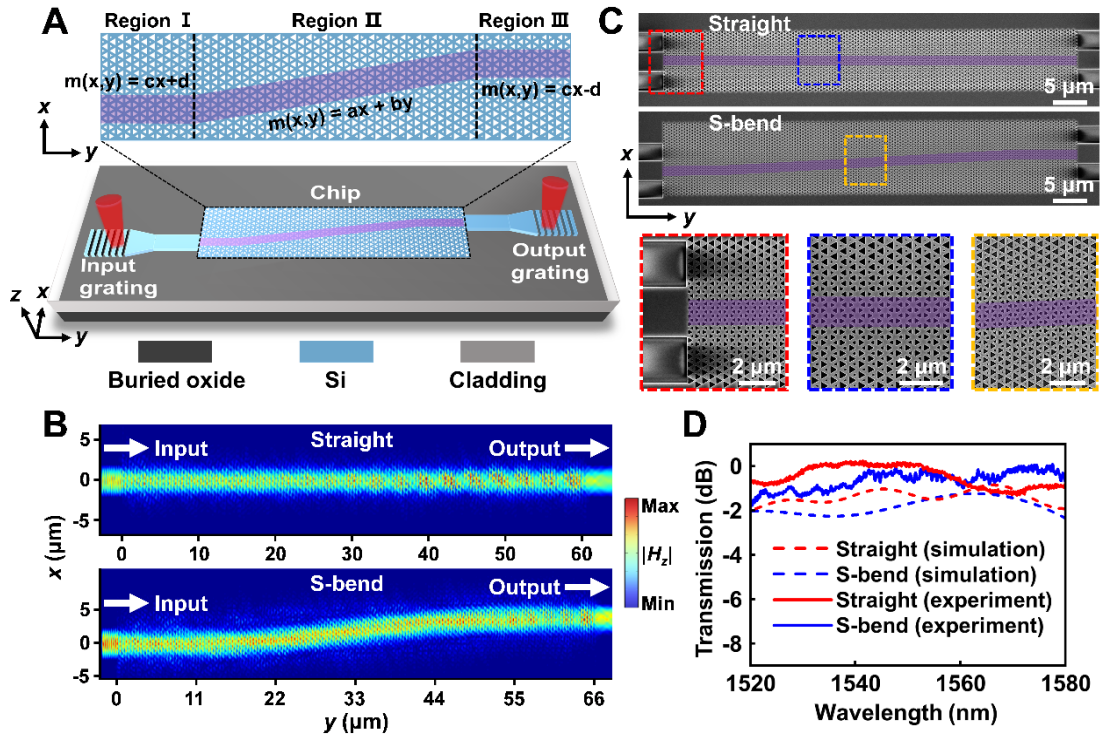

**Fig. 2. Demonstration of a straight waveguide and a S-bend based on PMFs.** (A) Schematic of the S-bend implemented on a SOI platform. The purple shaded area indicates the propagation path of light. Light is coupled into and out of the chip by grating couplers. (B) Simulated propagation profiles for the straight waveguide with $m(x) = ax$ (top) and the S-bend with a nonuniform PMF (bottom) at a central wavelength of 1550 nm. (C) SEM images of the fabricated devices. The purple shaded areas indicate the propagation paths of light. The areas encircled by the red, blue and orange dashed boxes are zoomed in and displayed on the bottom of the figure. (D) Simulated and measured transmission spectra of the straight waveguide and the S-bend.

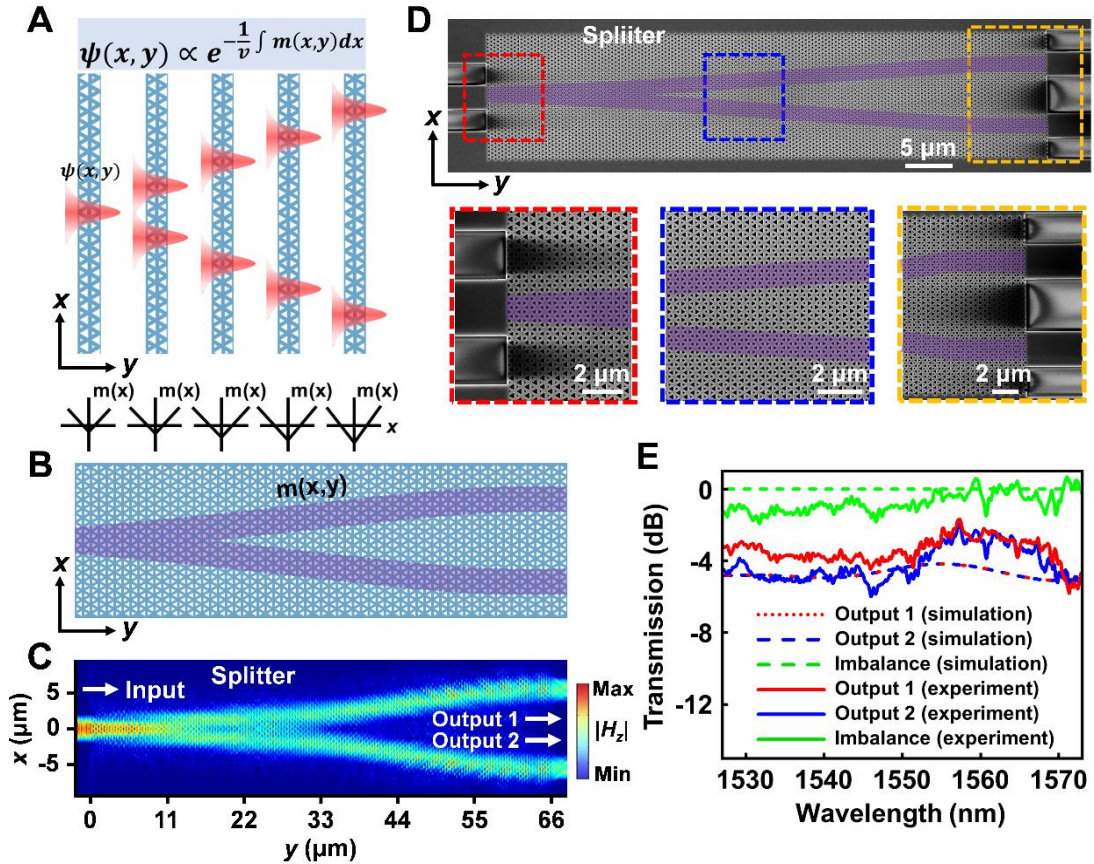

**Fig. 3. Demonstration of a 50:50 power splitter based on a PMF.** (A) Schematic illustration of the operation principle of the power splitter. (B) Schematic of the PhC structure of the splitter. The propagation paths of light are shaded in purple. (C) Simulated propagation profile for the PMF-based power splitter at a central wavelength of 1550 nm. (D) SEM images of the fabricated device. The purple shaded areas indicate the propagation paths of light. The areas encircled by the red, blue and orange dashed boxes are zoomed in and displayed on the bottom of the figure. (E) Simulated and measured transmission spectra of the power splitter.

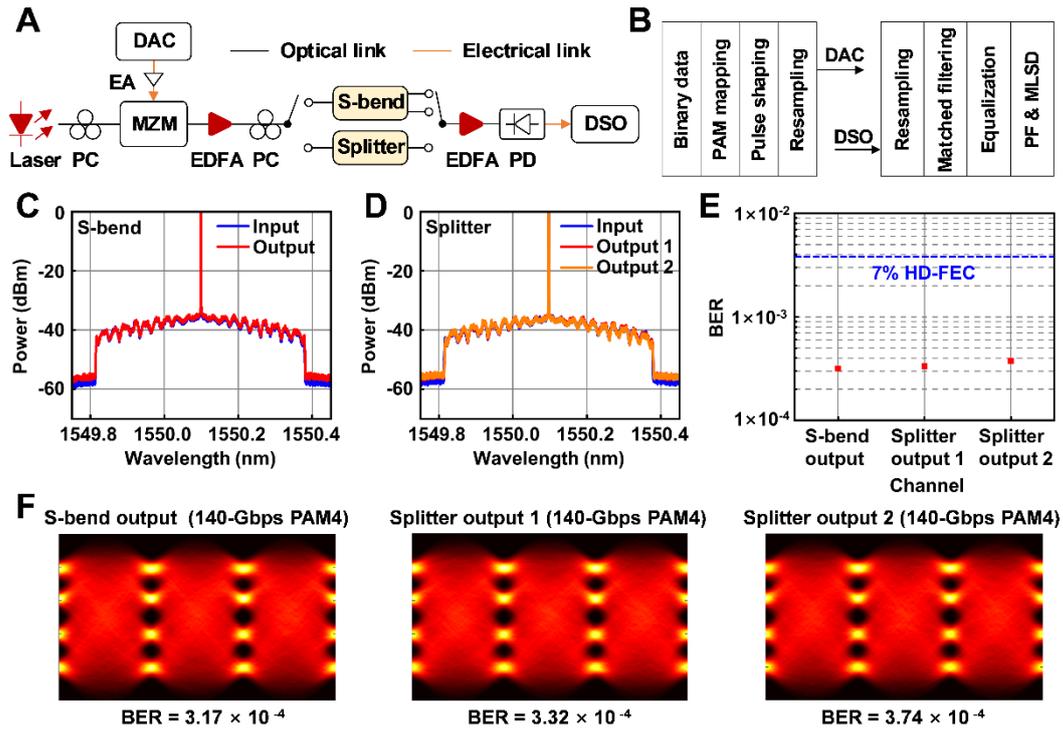

**Fig. 4. High-speed data transmission experiment based on the S-bend and the power splitter.** (A) Experimental setup for the data transmission of 140-Gb/s PAM-4 signals. The black and yellow lines represent optical and electrical links, respectively. PC, polarization controller; DAC, digital-to-analog converter; EA, electrical amplifier; MZM, Mach-Zehnder modulator; EDFA, erbium-doped fiber amplifier; PD, photodiode; DSO, digital storage oscilloscope. (B) DSP flow charts for the transceiver. (C and D) Measured optical spectra of the PAM-4 signals before and after passing (C) the S-bend and (D) the power splitter. (E and F) BERs (E) and recovered eye diagrams (F) of the PAM-4 signals for different channels of the S-bend and the splitter.